\documentstyle[aps,prl,twocolumn]{revtex}

\begin{document}
\draft
\title{Effects of transition metal doping in MgB$_2$ superconductor}
\author{Y. Moritomo}
\address{CIRSE, Nagoya University, Nagoya 464-8601, Japan}
\author{Sh. Xu}
\address{Department of Crystalline Materials Science, Nagoya University, Nagoya 464-8603, Japan}

\date{}
\maketitle

\begin{abstract}
Effects of chemical substitution of the divalent transition metals has been systematically investigated in Mg$_{1-x}M_x$B$_2$ ($x$ = 0.03; $M$ = Mn, Fe, Co, Ni and Zn).
Substitution of magnetic ions, {\it i.e.}, Mn$^{2+}$, Fe$^{2+}$, Co$^{2+}$ and Ni$^{2+}$, for the Mg$^{2+}$ ion suppresses $T_{\rm c}$: d$T_{\rm c}$/d$x$ is the largest (= - 159 K) in the Mn-doped sample.
We have found Zn-substitution increases $T_{\rm c}$ ($\Delta T_{\rm c} \approx$ 0.2 K at $x$ = 0.03), perhaps due to the enhanced density of state near the Fermi level.
\end{abstract}

\narrowtext
The recent discovery of the superconductivity in MgB$_2$ at $T_{\rm c}$ = 39 K\cite{akimitsu} has stimulated world wide excitement.
This is not only due to the simplicity in the chemical composition, the crystal structure and electronic structure, but also due to its potentiality for application.
Many researches are accumulating detailed information on the physical properties of the parent MgB$_2$.
On the other hand, the chemical substitution is one of the powerful approach not only to reveal the nature of the parent material, but also to enhance the material potentiality by elevating the transition temperature $T_{\rm c}$.
The parent MgB$_2$ has a hexagonal structure (AlB$_2$-type; space group $P6/mmm$)\cite{jones} with alternating B- and Mg-sheets.
The hexagonal network of the two-dimensional boron sheet governs the electronic structure near the Fermi level,\cite{band1,band2,band3} and hence is believed to be responsible for the superconductivity.
Band structural calculations showed that the electronic structure is rather three-dimensional, making a sharp contrast with the two-dimensional electronic structure of the graphite intercalation compounds. 
Another structural feature of MgB$_2$ is that a large number of isostructural compounds, such as, AlB$_2$, CrB$_2$ and MnB$_2$, exist.
These structural features have motivated attempts to substitute Mg with Li\cite{Li}, Al\cite{Al}, Zn\cite{Zn}, and Mn\cite{Mn} and B with C.\cite{C,C2}

In this Communication, we report substitution effects of the divalent transition metal ($M$ = Mn, Fe, Co, Ni and Zn) for the Mg$^{2+}$ ions on the superconductivity of MgB$_2$.
We have found substitution of the magnetic ions, {\it i.e.}, Mn$^{2+}$, Fe$^{2+}$, Co$^{2+}$ and Ni$^{2+}$, for the Mg$^{2+}$ ion suppresses $T_{\rm c}$, due to the interaction between the local spin $\vec{S}$ and the constituent electron $\vec{s}$ in the Cooper pair.
We have found the Zn-substitution increases $T_{\rm c}$ ($\Delta T_{\rm c} \approx$ 0.2 K at $x$ = 0.03), perhaps due to the enhanced density of state near the Fermi level.

The Mg$_{1-x}M_x$B$_2$ ($x$ = 0.03; $M$ = Mn, Fe, Co, Ni and Zn) samples were synthesized by heating a stoichometric mixture of amorphous boron (98 \%), magnesium powder (99.9 \%), Mn powder (99.9 \%), Fe powder (99.9 \%), Co powder (99 \%), Ni powder (99.9 \%), and Zn powder (99.9 \%) at 900 $^\circ$C for 2 hour.
The powders are place in a Ta foil and heated in a flow of Ar/H$_2$5\% gas.
The room temperature x-ray powder patterns are obtained in the 2 $\theta$ range of 10 - 100 degree with use of the Rigaku RINT2000/PC diffractometer.
The temperature dependence of magnetization $M$ was measured in a Quantum Design PPMS magnetometer under an applied field of 10 Oe.
The data were taken on heating after cooling down to the lowest temperature in zero field (ZFC).

First of all, let us show in Fig.\ref{fig1} the x-ray powder patterns of Mg$_{0.97}M_{0.03}$B$_2$ ($M$ = Mn, Fe, Co, Ni and Zn) at 300 K in the 2 $\theta$ range of 20 - 60 degree.
In all the compounds, sharp reflections from the AlB$_2$ structure are observed.
Small reflections indicated by open triangles (closed triangles) are due to MgB$_4$ (MgO) impurities.
In the case of the Ni-substituted sample, the powder patterns contain recognizable impurity peaks.
These peaks can be assigned to Ni-related impurities, such as, Ni$_4$B$_3$, MgNi$_2$ and so on, indicating that some part of the Ni atoms is wasted.
This situation is apparently similar to the case of Cu- and Ca-substitutions.
We tried to substitute Cu (Ca) for Mg, but the powder pattern contains considerable MgCu$_2$ (CaB$_6$) impurity peaks and $T_{\rm c}$ does not shift, indicating that non of the elements are substituted for Mg (not shown).
Nevertheless, not only the lattice constants but $T_{\rm c}$ shift in the Ni-doped sample, suggesting that a portion of the Ni atoms is substituted ({\it vide infra}).

The upper panels of Fig.\ref{fig2} show the magnified x-ray powder patterns around (a) (110) and (b) (002) reflections, whose peak positions reflect the in-plane and out-of-plane B-B lengths, respectively.
The lowest curve is the pattern for the non-doped MgB$_2$.
The (011) reflection shifts toward the low-angle side in the Co-, Ni- and Zn-substituted samples (see open triangles in Fig.\ref{fig2}(a)).
This indicates increase of the inter-plane B-B distance with chemical substitution.
Similarly, the (002) reflection shifts toward the low-angle side in the Fe-, Ni- and Zn-substituted samples (Fig.\ref{fig2}(b)), indicating increase of the inter-B-sheet distance.
In the Mn-substituted sample, the inter-B-sheet distance, rather, decreases.
To accurately determine the lattice constants, {\it i.e.}, $a$ and $c$, we have analyzed the powder pattern with the RIETAN-2000 program\cite{rietan} with AlB$_2$ structure ($P6/mmm$, No 191).
The impurity peaks were removed in the Rietveld refinement procedure.
The determined lattice constants, $a$ and $c$, are listed in Table~\ref{table}.
We plotted in the bottom panel (Fig.\ref{fig2}(c)) variation of the lattice constants, {\it i.e.}, $a$ and $c$ (see also Table~\ref{table}).
The Mn-doping effect on the lattice structure, that is decrease of $c$ and nearly constant $a$, is analogous to Mg$_{1-x}$Al$_x$B$_2$,\cite{Al} while the Co-doping effect, that is, increase of $c$ and nearly constant $c$, is similar to the Li-doped\cite{Li} and C-doped MgB$_2$.\cite{C}
The Zn-doping effects on the lattice structure, {\it i.e.}, elongation of both the lattice constant, is consistent with the work done by Kazakov {\it et al.}.\cite{Zn}

Figure~\ref{fig3} shows the temperature dependence of susceptibility $\chi$ for Mg$_{1-x}M_x$B$_2$ ($x$ = 0.03; $M$ = Mn, Fe, Co, Ni and Zn) together with the $\chi-T$ curve for the parent MgB$_2$.
The data were taken under applied filed of 10 Oe on heating after cooling in zero field (ZFC).
All the samples show well-defined one-step transitions as well as large shielding fraction before correction of demagnetization, indicating that the superconductivity is of bulk nature.
The transition temperature $T_{\rm c}$, defined by the intersection of the extrapolated lines below and above the transition, was listed in Table~\ref{table}.
In Fig.\ref{fig4} are plotted the relative change of $T_{\rm c}$ due to the 3 \% doping of the divalent transition metals.
Here, note that the actual doping level of the Ni-substituted sample is less than 3 \%, since the Ni-related impurities are observed in the x-ray powder pattern (see Fig.\ref{fig1}).
Inset shows a detailed doping dependence of $T_{\rm c}$ for Mg$_{1-x}$Mn$_x$B$_2$ (cited from Ref.\cite{Mn}).
$T_{\rm c}$ decreases linearly with Mn concentration $x$ at a rate of d$T_{\rm c}$/d$x$ = - 159 K.
If we assume a similar linear relation between $T_{\rm c}$ and $x$, d$T_{\rm c}$/d$x$ becomes - 83 K for the Co-substitution and - 13 K for the Fe-substitution.

The absolute magnitude of d$T_{\rm c}$/d$x$ of the Mn- and Co-doped samples are much larger as compared with the substitution of non-magnetic elements, such as Zn$^{2+}$ (present work) and Al$^{3+}$.\cite{Al}
Therefore, we ascribed the suppression of $T_{\rm c}$ to the interaction between the local spin $\vec{S}$ and the constituent electron $\vec{s}$ of the Cooper pair ($\sim$ - 2 $J \vec{S} \cdot \vec{s}$), not to the structural effect nor the electron doping effect.
Because the {\it local spin - electron interaction} tends to break the Cooper pairs, and to lower $T_{\rm c}$. 
Within this scenario, it is reasonable that the Mn$^{2+}$ ion, which has the largest local spin $S$ (= 5/2), shows the largest d$T_{\rm c}$/d$x$ (= - 169 K; see also Fig.\ref{fig4}).
Strangely enough, absolute magnitude of d$T_{\rm c}$/d$x$ (= - 13 K) for the Fe$^{2+}$ ion ($d^6$; $S$ = 2) is much smaller than the case of Mn$^{2+}$ ion ($d^5$; $S$ = 5/2)  and Co$^{2+}$ ion ($d^7$; $S$ = 3/2).
We think degree of the mixing between the $M$3$d$ level and the B2$p$ band is the key to solve this unexpected behavior, and hence elaborated band calculations are indispensable.

Finally, let us discuss on the Zn-substitution effect on $T_{\rm c}$.
Looking at Fig.\ref{fig3}, one may notice that $T_{\rm c}$ increases by $\approx$ 0.2 K by substitution of 3 \% Zn for the Mg.
Kazakov {\it et al.}.\cite{Zn} reported that $T_{\rm c}$ rather decreases by $\approx$ 0.5 K at $x$ =0.05 and by $\approx$ 0.2 K at 0.10.
These experimental observations suggest that there exists some optimal Zn concentration ($\sim$ 4 \%), beyond which $T_{\rm c}$ rather decreases.
Here, recall that the characteristic Zn-doping effect on the lattice structure is elongation of both the lattice constants (see Fig.\ref{fig2}).
The resultant increase of the in-plane and out-of-plane B-B distances reduces the one-electron bandwidth, and hence enhances the density of state (DOS) near the Fermi level $E_{\rm F}$.
This {\it structural effect} well explains increase of $T_{\rm c}$ in the Zn-doped sample.

In summary, chemical substitution effects on the superconductivity has been investigated in Mg$_{1-x}M_x$B$_2$ ($x$ = 0.03; $M$ = Mn, Fe, Co, Ni and Zn).
We have found suppression of $T_{\rm c}$ with substitution of magnetic ions, {\it i.e.}, Mn$^{2+}$, Fe$^{2+}$, Co$^{2+}$ and Ni$^{2+}$, for the Mg$^{2+}$ ion, even though these magnetic impurities occupy the Mg sites outside of the boron sheets.
This observation perhaps reflects three-dimensional electronic band structure of MgB$_2$.
A more elaborated and systematic investigation on the chemical substitution effects on $T_{\rm c}$ is necessary to reveal the nature of the MgB$_2$ superconductor. 

This work was supported by a Grant-in-Aid for Scientific Research from the Ministry of Education, Science, Sports and Culture.

\begin{table}
\caption{Lattice constants of Mg$_{1-x}M_x$B$_2$ ($x$ = 0.03; $M$ = Mn, Fe, Co, Ni and Zn) at 300 K. The transition temperatures $T_{\rm c}$ for the superconductivity were defined by the intersection of the extrapolated lines below and above the transition. Note that the doping-level is less than 3 \% in the Ni-doped sample (see text).}

\begin{tabular}{c|cc|c}
$M$&$a$ ($\AA$)&$c$ ($\AA$)&$T_{\rm c}$\\
\hline
Non-dope&3.08200(9)&3.52166(9)&38.2(1)\\
Mn&3.0827(4)&3.5188(3)&33.1(3)\\
Fe&3.0839(5)&3.5229(3)&37.8(1)\\
Co&3.0839(6)&3.5219(4)&35.7(2)\\
Ni&3.0841(8)&3.5243(5)&37.8(1)\\
Zn&3.0870(4)&3.5241(3)&38.4(1)\\
\end{tabular}

\label{table}
\end{table}

\begin{figure}
\caption{The x-ray powder pattern at 300 K for Mg$_{0.97}M_{0.03}$B$_2$ ($M$ = Mn, Fe, Co, Ni and Zn). Small reflections indicated by open triangles (closed triangles) are due to MgB$_4$ (MgO) impurities.}
\label{fig1}
\end{figure}

\begin{figure}
\caption{(a) Magnified powder patterns around (110) reflection for Mg$_{0.97}M_{0.03}$B$_2$ ($M$ = Mn, Fe, Co, Ni and Zn). The lowest curve is the pattern for the non-doped MgB$_2$. (b) Magnified powder patterns around (002) reflection for Mg$_{0.97}M_{0.03}$B$_2$. The lowest curve is the pattern for the non-doped MgB$_2$. (c) Variation of the lattice constants, {\it i.e.}, $a$ and $c$, for Mg$_{0.97}M_{0.03}$B$_2$.}
\label{fig2}
\end{figure}

\begin{figure}
\caption{Temperature dependence of susceptibility $\chi$ for Mg$_{0.97}M_{0.03}$B$_2$ ($M$ = Mn, Fe, Co, Ni and Zn). The thick curve represent the $\chi-T$ curve for the non-doped MgB$_2$. The data were taken under applied filed of 10 Oe on heating after cooling in zero field (ZFC). Downward arrows indicate the transition temperatures $T_{\rm c}$. Note that the doping-level is less than 3 \% in the Ni-doped sample (see text).} 
\label{fig3}
\end{figure}

\begin{figure}
\caption{Variation of $T_{\rm c}$ for Mg$_{0.97}M_{0.03}$B$_2$ ($M$ = Mn, Fe, Co, Ni and Zn). Inset shows doping dependence of $T_{\rm c}$ for Mg$_{1-x}$Mn$_x$B$_2$ (cited from Ref.9).}
\label{fig4}
\end{figure}


\begin{references}

\bibitem{akimitsu} J. Nagamatsu, N. Nakagawa, T. Muranaka, Y. Zenitani and J. Akimitsu, Nature, {\bf 410} (2001) 63.
\bibitem{jones} M. E. Jones and R. E. Marsh, J. Am. Chem. Soc., {\bf 76} (1954) 1434.
\bibitem{band1} S. Suzuki, S. Higai and K. Nakao, J. Phys. Soc. Jpn., submitted.
\bibitem{band2} G. Satta, G. Profeta, F. Bernardini, A. Continenza and S. Massidda, cod-mat/0102358
\bibitem{band3} J. E. Hirsch, {\it et al}, cond-mat/0102115
\bibitem{Li} Y. G. Zhao, {\it et al}, cond-mat/0103077  .
\bibitem{Al} J. S. Slusky, {\it et al}, Nature, {\bf 410} (2001) 243.
\bibitem{Zn} S. M. Kazakov and M. Angst and J. Karpinski, cond-mat/0103350.
\bibitem{Mn} Sh. Xu, Y. Moritomo, K. Kato and A. nakamura, cond-mat/0104534.
\bibitem{C} T. Takenobu, T. Ito, Dam H. Chi, K. Prassides and Y. Iwasa, cond-mat/0104086.
\bibitem{C2} M. Paranthaman, J. R. Thompson, D. K. Christen, cond-mat/0103241.
\bibitem{rietan} F. Izumi; {\it The Rietveld Method}, ed. R. A. Young (Oxford 
University Press ,Oxford, 1993, Chap.13.
%\bibitem{DOS1} A. L. Ivanovskii and N. I. Medvedeva, Russian J. Inorg. Chem. {\bf 45} (2000) 1234.
%\bibitem{DOS2} N. I. Medvedeva, A. L. Ivanovskii, J. E. Medvedeva and A. J. Freeman, cond-mat/0103157. 
\end{references}
\end{document}